\documentclass{PoS}
\usepackage{amsmath,amssymb,amsfonts}
\usepackage{slashed}
\usepackage{bbm}
\usepackage{cite}
\usepackage[utf8]{inputenc}
\usepackage[english]{babel}

\title{What we can learn from two-dimensional QCD-like theories at finite density}
\ShortTitle{Two-dimensional QCD-like theories}
\author{\speaker{Bj\"orn H. Wellegehausen}\thanks{Discussions and collaboration with Axel Maas and Andreas Wipf are gratefully acknowledged. 
This work was supported by the Helmholtz International Center for FAIR within the LOEWE initiative of the State of Hesse.
Simulations were performed on the LOEWE-CSC at the University of Frankfurt.}\\%
        Justus-Liebig-University Giessen\\
        E-mail: \email{bjoern.wellegehausen@theo.physik.uni-giessen.de}}
\author{Lorenz von Smekal\\%
        Justus-Liebig-University Giessen and TU Darmstadt\\
        E-mail: \email{lorenz.smekal@physik.tu-darmstadt.de}}

\abstract{
We study generic properties of strongly interacting matter at finite density as relevant to heavy-ion collisions at moderate beam energies or the physics of neutron stars and their mergers. 
Because of the fermion-sign problem in lattice QCD, here we simulate QCD-like theories without this problem at finite density. 
These theories (two-color QCD, G2-QCD, or adjoint QCD) typically contain bosonic baryons, for example diquarks, or other more exotic states of matter. 
It is therefore important to understand the effects of such bosonic matter and disentangle them from fermionic baryons where they exist to draw conclusions for QCD. 
Simulations of these theories, for instance G2-QCD, reveal an interesting and rich phase diagram at zero temperature. 
Many open questions arise, partly due to the lack of high precision or large volume/continuum data. 
This is the reason why we study two-dimensional QCD-like theories. In this contribution we shall discuss differences between QCD-like theories at baryon chemical and isospin chemical potential. 
Furthermore we present simulation results on the phase diagram and spectroscopy at finite density for G2- and two-color-QCD and compare it to free lattice fermions.
}

\FullConference{34th annual International Symposium on Lattice Field Theory\\
		24-30 July 2016\\
		University of Southampton, UK}
\usepackage{etex}
\usepackage{graphicx}
\usepackage{bbm,bm}
\usepackage{mathtools}
\usepackage{color,array,tabularx}
\usepackage{stackrel}
\usepackage[small,bf]{caption}
\usepackage{pst-all}
\usepackage{psfrag}
\usepackage{pstricks,pst-grad,color,shadow}
\usepackage{tikz}
\usetikzlibrary{shapes,arrows,backgrounds}
\usepackage{soul}
\usepackage{placeins}
\usepackage{latexsym}
\usepackage{keyval}
\usepackage{simplewick}
\usepackage{epigraph}
\usepackage{amsthm}

\definecolor{darkgreen}{rgb}{0,0.73,0}

\newcommand{\id}{\mathbbm{1}}

\newcommand{\trnsp}{\mathsf{T}}

\newcommand{\ii}{\mathrm{i}}

\graphicspath{{./}{./plots/}{./publPlots/}}

\begin{document}

\section{Introduction}
\noindent

Although there is a lot of progress in understanding the QCD phase diagram from lattice simulations of QCD at finite baryon density (complex langevin, etc\dots) \cite{Aarts:2016qrv} and functional methods like the FRG method \cite{Braun:2009gm}, 
QCD-like theories still play an important role to understand various aspects of the phase diagram as for instance chiral symmetry breaking, deconfinement, diquark condensation or the transition from the vacuum to nuclear matter.
QCD-like theories replace the fundamental $SU(3)$ fermions by fermions in a different representation or gauge group in order to have a positive fermion determinant \cite{Kogut:2000ek}. 
The most important QCD-like theories are adjoint QCD \cite{Bilgici:2009jy}, two-color QCD, that has been studied with much effort in the last years \cite{Cotter:2012mb,Holicki:2017psk}, and $G_2$-QCD \cite{Wellegehausen:2015iea}. All of them share different aspects with QCD, but also have their shortcommings.
While two-color QCD contains only bosonic baryons, and is therefore an ideal setup to study diquark condensation, it is not suitable to investigate the nuclear matter transition of QCD. $G_2$-QCD features bosonic as well as different
kinds of fermionic baryons, but it is difficult to disentangle their effects in the phase diagram. 
Furthermore from a computational point of view it is much more expensive. This is one reason why we compare high precision simulations of two-color QCD and $G_2$-QCD in $1+1$ dimensions at finite baryon density. 
Although there is no spontaneous symmetry breaking in two dimensions, in a finite volume these theories
resemble their four-dimensional versions quite well. 
Another important question is whether the fermion sign problem of QCD is important for the physics at non-vanishing baryon density
or whether these QCD-like theories are more similar to QCD with isospin chemical potential.

A sufficient condition for a real fermion determinant is to demand an additional (anti-) unitary symmetry for the dirac operator with real baryon chemical potential $D(\mu)$ \cite{Kogut:2000ek},
\begin{equation}
 T D(\mu)= D^*(\mu) T\,,\quad T^*T=\pm \id\,,\quad T^\dagger T=\id.
\end{equation}
Under this condition, the fermion determinant for two mass degenerated fermion flavours is always positive,
\begin{equation}
 \det D(\mu) \det D(\mu)=\det D(\mu) \det D^*(\mu)=\det \left(D(\mu) D^\dagger(\mu)\right)\geq 0
\end{equation}
and standard Monte-Carlo simulations at finite baryon chemical potential are applicable to investigate the phase diagram at finite density and temperature non-perturbatively.
For $T^*T=-\id$, the fermion determinant for a single flavour is already positive, and simulations with a single flavour are possible, for instance four-dimensional $G_2$-QCD.
For QCD-like theories with $\gamma_5$-hermiticity, i.e. $D(\mu)=\gamma_5 D^\dagger(-\mu) \gamma_5$, the fermion determinant at isospin chemical potential is also positive,
\begin{equation}
 \det D(\mu) \det D(-\mu)=\det D(\mu) \det D^\dagger(\mu)=\det \left(D(\mu) D^\dagger(\mu)\right)\geq 0
\end{equation}
such that for theories with both symmetries the two-flavour partition functions for baryon and isospin chemical potential are the same, i.e.  $Z(\mu_\text{B})=Z(\mu_\text{I})$.
Therefore the phase diagrams at isospin and baryon chemical potential are also identical under the following mapping for the up- and down-quark content $(u,d)$ in observables:
\begin{equation}
 (u,d) \mapsto (u,T \gamma_5 \bar{d}^{\,\trnsp}).
\end{equation}
This symmetry changes the fermion number of an operator, for instance it relates mesons and diquarks, and is not allowed in QCD where it breaks gauge invariance. 
This observation suggests that these theories might have more in common with isospin QCD than with QCD at baryon chemical potential.
In this paper we consider two-color QCD and $G_2$-QCD.

Two-color QCD, QCD with fermions in the fundamental representation of $SU(2)$, contains only bound states with an even quark number. 
These are either mesons or bosonic baryons like for instance diquarks with fermion
number two. In contrast to QCD, fermionic baryons are forbidden by gauge invariance. The antiunitary symmetry is related to the operator
$T=C \gamma_5 \times \sigma_2$ with charge conjugation matrix $C$ and $\sigma_2$ acting on colour space. The mapping between the lightest bilinear bound states is shown in Table \ref{su2map}(left).
\begin{table}[htb]
\begin{tabular}{|c|c|c|c|c|}
\hline $n_\text{q}$ & Particle & $d \leftrightarrow T\, \gamma_5 \, \bar{d}^\trnsp$ & Particle & $n_\text{q}$ \\
\hline
\hline 0  & $\eta$&  $\leftrightarrow$ & $\eta$ & 0 \\
\hline 0  & $f$&  $\leftrightarrow$ & $f$ & 0 \\
\hline 0  & $\pi_0$&  $\leftrightarrow$ & $\pi_0$ & 0 \\
\hline
\hline 0  & $\pi_{\pm}$&  $\leftrightarrow$ & $d_{\pm}^{+}$ & 2 \\
\hline 0  & $a_{\pm}$&  $\leftrightarrow$ & $d_{\pm}^{-}$ & 2 \\
\hline
\end{tabular}\hskip5mm
\begin{tabular}{|c|c|c|c|c|}
\hline $n_\text{q}$ & Particle & $d \leftrightarrow T \gamma_5 \, \bar{d}^\trnsp$ & Particle & $n_\text{q}$ \\
\hline
\hline 1 & $H$&  $\leftrightarrow$ & $H$ & 1 \\
\hline 1 & $\tilde{N}$&  $\leftrightarrow$ & $N$ & 3\\
\hline 1 & $\tilde{\Delta}^{++,+,-}$&  $\leftrightarrow$ & $\tilde{\Delta}^{++,+,-}$ & 1\\
\hline 1 & $\tilde{\Delta}^0$&  $\leftrightarrow$ & $\Delta^0$ & 3\\
\hline 3 & $\Delta^{++,+,-}$&  $\leftrightarrow$ & $\Delta^{++,+,-}$ & 3\\
\hline
\end{tabular}
\caption{Mapping between bilinear bound states (left) and fermionic baryons (right) under the (anti-)unitary symmetry (charge conjugation).}
\label{su2map}
\end{table}

In $G_2$-QCD we replace the fundamental $SU(3)$ fermions by fermions in the fundamental $7$-dimensional representation of the execptional Lie group $G_2$. In this case the operator $T$ is given by $T=C \gamma_5 \times \id$. In addition to the bound states of two-color QCD and QCD, it also contains
quark-gluon hybrids ($H$) and quark-meson bound states ($\tilde{N}$, $\tilde{\Delta}$), see \cite{Wellegehausen:2015iea}. Their relation under down-quark charge conjugation is shown in Table \ref{su2map}(right). 
Coupled to a Higgs field in the fundamental representation, this theory reduces to isospin QCD for a non-vanishing Higgs field vacuum expectation value, indicating that the theory shares many aspects with isospin QCD.

\section{$G_2$-QCD in 4 dimensions}
\begin{figure}[htb]
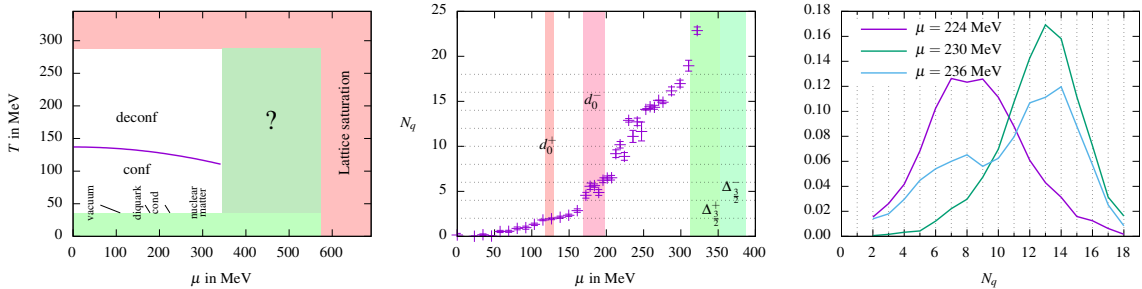

    \scalebox{0.71}{\input{plots/Phasediagram_Light_}}
     \scalebox{0.71}{\input{plots/QND_Light_NT16_}}
     \scalebox{0.71}{\input{plots/qndHist_}}
  \caption{\textbf{Left}: Phase diagram of four-dimensional $G_2$-QCD on a $N_t\times 8^3$ lattice at $\beta/N_\text{c}=0.96$ and $\kappa=0.159$. \textbf{Center}: Quark number compared to diquark and nucleon masses $m/n_q$. \textbf{Right}: Histogram of the quark number around the first order transition.} 
     \label{PDG2D4}
   \end{figure}

\noindent
The phase diagram of $G_2$-QCD on a rather small $N_t\times8^3$ lattice is shown in Figure \ref{PDG2D4} (left).
Confinement and deconfinement phase are seperated by a crossover at $T_\text{c}(\mu=0)\approx 137$ MeV. With increasing chemical potential, the transition shifts to smaller temperatures as expected from
recent QCD calculations \cite{Cea:2015cya}. The quark number $N_q$ for our smallest temperature $T\approx 36$ MeV (Fig.~\ref{PDG2D4}, center) shows an onset transition to $N_q=2$ close 
to half of the mass of the lightest diquark followed by various transitions at larger values of chemical potential. Most of these transition lead to a plateau in the quark number that can possibly be mapped
to an appropriate distribution of baryons on the finite number of lattice sites. The first two transitions at $\mu\approx 110$ MeV and $\mu\approx 170$ MeV are related to diquarks with positive and negative parity and the quark number seems to be 
a continuous function. At $\mu \approx 230$ MeV we observe probably a first order transition in the quark number which can be seen as a jump in the density and a phase coexistence in the corresponding
histograms of the quark number (Fig.~\ref{PDG2D4}, right). Unfortunately within given statistical errors, we cannot decide whether the jump in the quark number is even or odd and therefore related to
bosonic or fermionic baryons. This is the reason why we investigate two-dimensional QCD-like theories where we can perform high-precision simulations that may help to understand the behaviour in four dimensions.

\section{Free lattice fermions}
\noindent
First we show some results for free lattice Wilson fermions in two dimensions. In order to mimic diquark bound states, we project the partition function for free fermions $Z(\mu)$ onto an ensemble with
even quark number,
\begin{equation}
 Z_\text{even}(\mu)=\frac{1}{2}\left(Z(\mu)+Z(\mu - \ii \pi T)\right).
\end{equation}
This corresponds to the sum of ensembles with periodic and anti-periodic boundary conditions in temporal direction. 
\begin{figure}[htb]
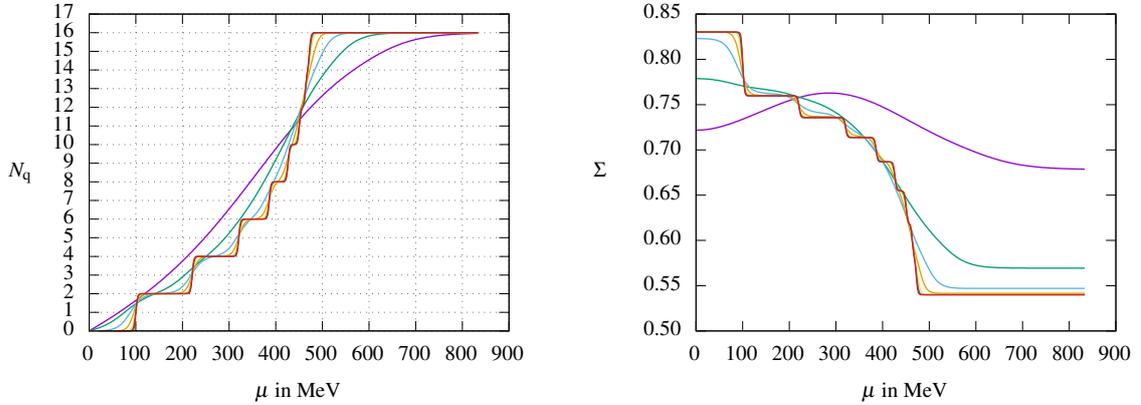

    \scalebox{1}{\input{plots/FreeDensity_}}\hskip10mm
    \scalebox{1}{\input{plots/FreeCondensate_}}
  \caption{Quark number (left) and chiral condensate (right) for free Wilson fermions and different temporal lattice sizes. The $\kappa$ parameter is tuned such that the first onset transition takes place at $\mu=100$ MeV.} 
     \label{FreeFermions}
   \end{figure}
The quark number and the chiral condensate, shown in Figure \ref{FreeFermions}, are given by derivatives with respect to $\mu$ and $m$,
\begin{equation}
 N_q=T\,\frac{d \ln Z_\text{even}(\mu)}{d \mu}\,,\quad \Sigma=\frac{1}{V}\frac{d \ln Z_\text{even}(\mu)}{d m}.
\end{equation}
In the zero temperature limit, the quark number increases by steps of two until the lattice is completely filled with $8$ \emph{diquarks} with increasing relative momentum. 
The length of the plateaus is related to the spatial size of the lattice such that in the infinite volume limit the quark number is continuously increasing. Whenever a \emph{diquark} is put on the lattice,
the chiral condensate decreases.

\section{Two-Color QCD in two dimensions}
\begin{figure}[htb]
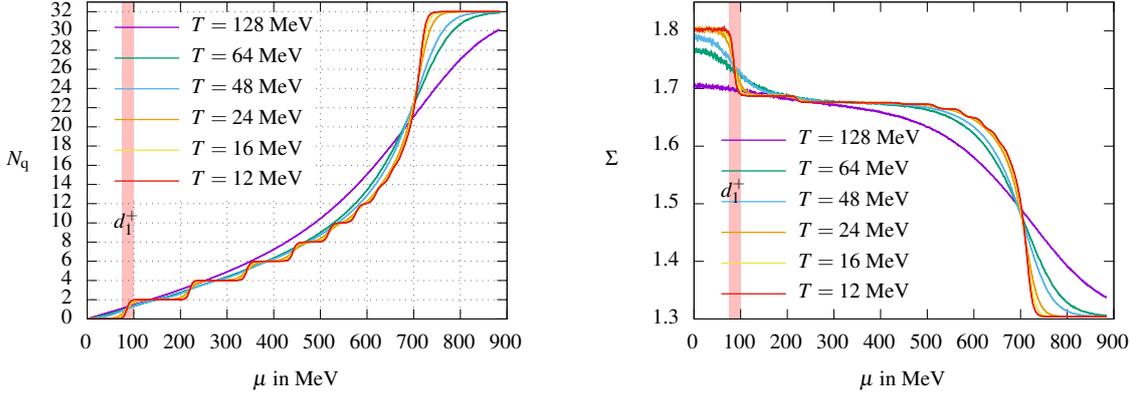

    \scalebox{1}{\input{plots/A1DensityLines_}}\hskip10mm
    \scalebox{1}{\input{plots/A1ChiralLines_}}
  \caption{Quark number per flavour (left) and chiral condensate (right) for two-color QCD on a $N_t \times 16$ lattice.} 
     \label{twocolor}
   \end{figure}

\begin{figure}[htb]
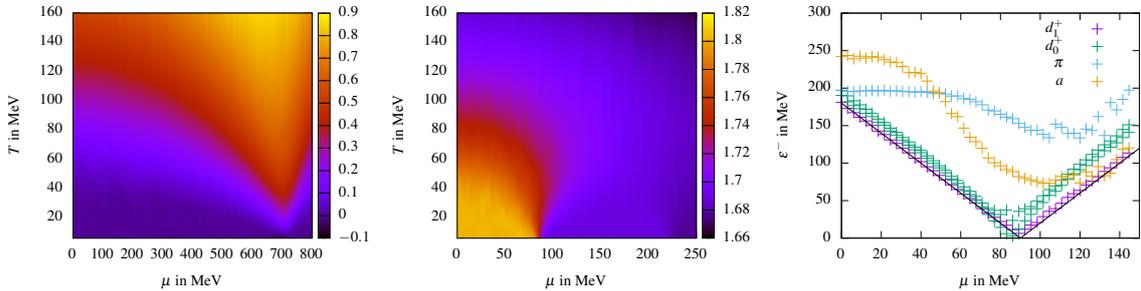

    \scalebox{0.71}{\input{plots/A1PDPol_}}
    \scalebox{0.71}{\input{plots/A1PDChiral_}}
    \scalebox{0.71}{\input{plots/A1Spectroscopy_}}
  \caption{Polyakov loop (left) and chiral condensate (center) for two-color QCD on a $N_t \times 16$ lattice and diquark and meson masses in two-color QCD at finite chemical potential (right)} 
     \label{twocolorPD}
   \end{figure}

\noindent
The simulations for two-color QCD have been performed on a $N_t \times 16$ lattice with $N_t=2 \dots 128$ at fixed gauge coupling $\beta/N_\text{c}=1.9$ and hopping parameter $\kappa=0.273$. Physical units are set by the pion mass
$m_\pi=200$ MeV at $N_t=32$, leading to a lattice spacing of $a=0.26(4)$ fm. This corresponds to temperatures between $T\approx6$ MeV and $T\approx385$ MeV. 
For this ensemble, the lightest bound state is the positive parity vector diquark with mass $m_{d_1^+}\approx177$ MeV, followed by the scalar diquark with mass $m_{d_0^+}=200$ MeV and the $a$-meson with mass $m_a\approx 254$ MeV. 
The results for the quark number and chiral condensate are shown in Figure \ref{twocolor}.
   With decreasing temperature the onset transition to diquark matter at half of the mass of the lightest baryon becomes more pronounced. The chiral condensate decreases significantly and we expect a diquark condensate to be formed. The first onset transition is
   followed by various transitions where the quark number always increases by two. In contrast to free fermions, we can put diquarks with different relative momenta or different kinds of diquarks on the lattice. All these transitions
   show also up as a small drop in the chiral condensate. When the lattice saturates at $N_\text{q}=N_\text{s} \, N_\text{c}=32$, the chiral condensate decreases to its quenched value. The phase diagram for the Polyakov loop and the chiral
   condensate are shown in Figure \ref{twocolorPD} (left and center).
   Qualitatively, they agree with the phase diagram obtained in four dimensions, but at very low temperatures, the Polyakov loop is always zero in contrast to results with Wilson fermions at the smallest temperatures in four dimensions \cite{Holicki:2017psk}. 
   In simulations with staggered fermions, the Polyakov loop is always zero below $T_\text{c}(\mu)$ \cite{Holicki:2017psk}, indicating that discretization effects already become important well below half filling.
   At finite density correlation functions of bound states with fermion number $n_q$ can be fitted to
   \begin{equation}
    C(\mu,n_q) \sim a e^{-\epsilon^-(\mu,n_q)}+b e^{\epsilon^+(\mu,n_q)} \quad \text{with} \quad \epsilon^\pm(\mu,n_q)=m(\mu)\,\pm\, n_q \mu .
   \end{equation}
   For the lightest diquarks and mesons, the function $\epsilon^-$ is shown in Figure \ref{twocolorPD} (right).
   The correlation function is fitted with 2,3 or 4 exponential factors for the ground- and excited states at $N_t=32$.
   As expected, below the critical $\mu_\text{c}\approx 90$ MeV the diquark masses do not depend on the chemical potential and $\epsilon^-$ is a linear function with slope $n_q=1$. Close to the onset, the scalar diquark tends to become lighter
   than the vector diquark and the pion mass decreases. Above the onset, the meson masses increase as $m\sim \mu$ while the vector diquark mass vanishes. The scalar diquark becomes again heavier than the vector diquark, but its mass does almost not depend on chemical potential.
   The mass of the $a$-meson decreases already before the critical $\mu_\text{c}$, indicating that finite temperature effects might be important (a similar behaviour is seen in the chiral condensate at $N_t=32$).

\section{$G_2$-QCD in two dimensions}
\begin{figure}[htb]
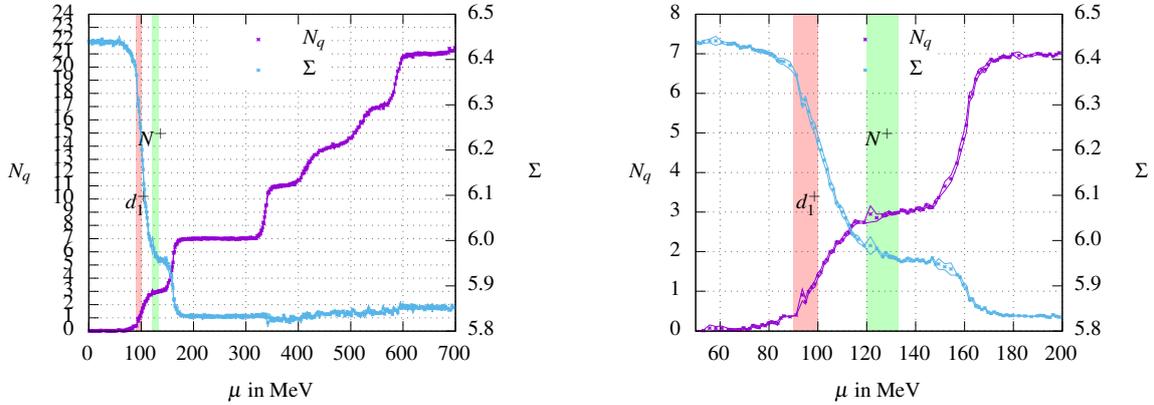

    \scalebox{1}{\input{plots/G2DensityLines_}}\hskip10mm
     \scalebox{1}{\input{plots/G2DensityLinesDetail_}}
  \caption{Quark number per flavour and chiral condensate for $G_2$-QCD on a $64 \times 16$ lattice} 
     \label{g2qcd}
   \end{figure}

\begin{figure}[htb]
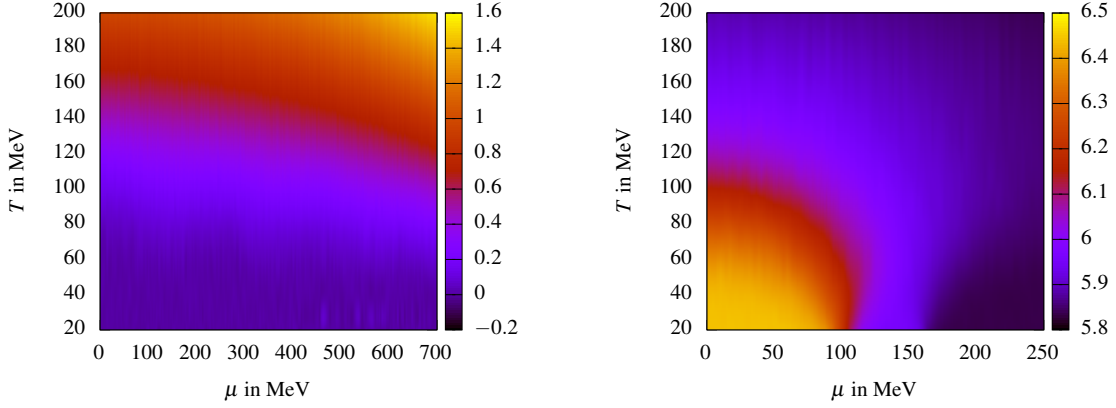

    \scalebox{1}{\input{plots/G2PDPol_}}\hskip10mm
    \scalebox{1}{\input{plots/G2PDChiral_}}
  \caption{Polyakov loop (left) and chiral condensate (right) for $G_2$-QCD on a $N_t \times 16$ lattice.} 
     \label{g2qcdPD}
   \end{figure}

\noindent
The simulations for $G_2$-QCD are done on a $N_t \times 16$ lattice at $\beta/N_\text{c}=3.1$ and $\kappa=0.275$. Again, the physical scale is set by the pion mass $m_\pi=200$ MeV, leading to a lattice spacing of $a\approx 0.16$ fm and temperatures
from $T\approx 20 \dots 633$ MeV. The lightest particle is again the vector diquark with mass $m_{d_1^+}\approx 194$ MeV, followed by $m_{d_0^+}=m_a\approx 262$ MeV, the nucleon masses $m_{N^+}\approx 380$ MeV, $m_{N^-}\approx 506$ MeV 
and the hybrid mass $m_H\approx 440$ MeV. The results for the quark number and the chiral condensate at the lowest temperature $T\approx32$ MeV are shown in Figure \ref{g2qcd} (left).
Similar to two-color QCD, various transition show up in the quark number
and the condensate, the first close to half of the vector diquark mass, but the increase in quark number is not always two. The lattice is filled with combinations of diquarks, nucleons and hybrids and with the present data it is impossible to understand every transition. 
At small densities, two plateaus in the quark number show up, the first at $N_q=3$ and the second at $N_q=7$ (Figure \ref{g2qcd}, right).
A plateau at $N_q=2$ is not visible, maybe because the transition to diquark matter at $\mu_\text{c}\approx 95$ MeV is influenced by the transition to nuclear matter. We expect that a plateau at $N_q=2$ will show up at smaller temperatures.
Another explanation might be that after the first onset the nucleon mass depends on chemical potential. First results here indicate indeed, that the nuclear mass decreases with chemical potential close to  $\mu_\text{c}$.
The phase diagram, Figure \ref{g2qcdPD}, looks very similar to the two-color phase diagram.

\section{Conclusions}\label{sconclusions}
\noindent
High precision simulations of QCD-like theories in two dimensions at finite density show many features that are also expected in four dimensions.
It turns out, that the phase diagrams of $G_2$-QCD and two-color QCD are very similar. In order to disentangle effects from bosonic and fermionic bound states in the density, simulations at very low temperatures 
(much lower than the achieved temperatures in four dimensional $G_2$-QCD) are necessary. Whether their phase diagrams are similar to QCD at baryon chemical potential is still unclear, at least these theories share important feature with
isospin QCD.

\end{document}